\documentstyle[prc,aps,epsfig,twocolumn,eqsecnum,subfigure]{revtex}

\catcode`\@=11

\def\maketitle2{\par 
\begingroup
\let\cite\@bylinecite
\def\thefootnote{\fnsymbol{footnote}}%
\twocolumn[\@maketitle2\vskip2pc]%
\thispagestyle{plain}\@thanks
\endgroup
\def\thefootnote{\arabic{footnote}}%
\setcounter{footnote}{0}%
\let\maketitle2\relax \let\@maketitle2\relax
\let\@thanks\relax \let\@authoraddress\relax \let\@title\relax
\let\@date\relax \let\thanks\relax \let\@abstract\relax 
\let\@pacs\relax}

\def\abstract#1{\gdef\@abstract{{\par 
\bgroup
\ifdim\prevdepth=-1000pt \prevdepth0pt\fi
\hsize\columnwidth
\dimen0=-\prevdepth \advance\dimen0 by17.5pt \nointerlineskip
\small\vrule width 0pt height\dimen0 \relax}{~~}#1\egroup}}

\def\pacs#1{\gdef\@pacs{{\par 
\bgroup
\hsize\columnwidth \parindent0pt
\ifdim\prevdepth=-1000pt \prevdepth0pt\fi
\dimen0=-\prevdepth \advance\dimen0 by20pt\nointerlineskip
\egroup} PACS numbers:~#1}}

\def\@maketitle2{
\@preprint
\@title
\ifdim\prevdepth=-1000pt \prevdepth0pt\fi
\@authoraddress
\@date
\begin{list}{}{\leftmargin=0.10753\textwidth \rightmargin=\leftmargin
\itemsep=1pc\partopsep=-1pc}
\item\@abstract
\item\@pacs
\end{list}
}

\catcode`\@=12

\begin{document}
\draft

\title{Ground state correlations and mean-field in $^{16}$O}
\author{Jochen~H.~Heisenberg\thanks{e-mail:Jochen.Heisenberg@unh.edu} 
        and Bogdan~Mihaila\thanks{e-mail:Bogdan.Mihaila@unh.edu} }
\address{Department of Physics, University of New Hampshire, \\
         Durham, New Hampshire 03824}

\preprint{}
\date{\today}

\abstract
{
   We use the coupled cluster expansion ($\exp(S)$ method) to generate the 
   complete ground state correlations due to the $NN$ interaction. 
   Part of this procedure is the calculation of the two-body ${\mathbf G}$ 
   matrix inside the nucleus in which it is being used. 
   This formalism is being applied to $^{16}$O in a configuration space 
   of 50 $\hbar\omega$. 
   The resulting ground state wave function is used to calculate the binding 
   energy and one- and two-body densities for the ground state of~$^{16}$O.
}

\maketitle2 
\narrowtext

\section{Introduction}

In the last thirty years electron scattering from nuclei has provided a
wealth of information mapping out nuclear ground state charge 
densities \cite{ref:Cavedon},
providing precise transition charge and current densities for the
excitation of single particle states \cite{ref:SchwentkerETal} and for 
collective states \cite{ref:GoutteETal}. The
measurement of ground state magnetization densities and the excitation
of high multipolarity magnetic excitations, or the single particle
knockout reaction to discrete states all have in some way supported
the mean-field approach as the lowest order in the description of
nuclear structure.

The confirmation of the mean-field approach, however, were more
qualitative in nature than quantitative. 
The form factors e.g. for the
excitation of the high spin single particle states in $^{208}$Pb
\cite{ref:LichtenstadtETal},
were described extremely well in shape by the mean-field wave functions,
however, the predicted strength was too big by a factor of two. The knockout
reactions again were in good agreement with the shapes predicted by the
mean-field wave functions but the strength was off by again roughly a
factor of two~\cite{ref:LeuschnerETal}. The general conclusion was that 
correlations are
important in the calculation of observables. They do not change the
shape of the wave functions but they modify the strength due to
deoccupation of orbits below the Fermi surface and partial occupation
of the orbits above the Fermi surface \cite{ref:PandharipandeETal}. 
This was confirmed by
(e,e$'$p) experiments in which particles from orbits above the Fermi
level were knocked out \cite{ref:LeuschnerETal}.
Thus, to do justice to the
accuracy of the electromagnetic probe, we can no longer be satisfied
with the mean-field approach but have to take into account the
correlations largely due to the hard repulsive core of the
nucleon-nucleon interaction. While we put particular emphasis on the
effects of correlations on the results of electron scattering
experiments this is less of an issue in this paper dealing with the
ground state, but will be more clearly pointed out in subsequent papers
dealing with excited states of these nuclei.

There are different ways to account for correlations. One way is to
introduce correlation functions in the many body wave function in
real space. This has been quite successful for
small nuclei~\cite{ref:Pudlineretal_2} and has resulted in reasonable 
descriptions of the $^{16}$O nucleus~\cite{ref:PieperETal}. 
A different approach is to add in configuration space to
the uncorrelated ground state multi-particle multi-hole configurations.
Both approaches can be related to each other
even though it seems harder using correlation functions to satisfy
the Pauli principle. In configuration space this appears to be more
transparent. In our treatment we follow closely the formulation of
the Bochum group \cite{ref:KummelETal}. However, we solve the equations 
entirely in configuration space. Furthermore, we truncate in different ways
where the significance of terms becomes more transparent

\section{The uncorrelated ground state}

We assume an orthonormal set of single particle  wave
functions exists. These wave functions are solutions to the 
single particle Hamiltonian given by the 
Schr\"odinger equation in the mean-field. This mean-field will be
generated in an iterative procedure. The single particle
wave functions are expanded in a set of orthonormal functions such as
harmonic oscillator functions, Bessel functions or others. Thus each
orbit is represented by a set of expansion coefficients and
a single particle energy. An uncorrelated ground state can be
constructed as a single Slater determinant which includes all the occupied
orbits and is written as $| 0\rangle$, our new vacuum state.

\section{Variational calculations for the correlated ground state}

We use the formalism of second quantisation. As such we follow the
notation of the textbook by deShalit and Feshbach \cite{ref:deShalitFesh}.
In the second quantisation representation we write the Hamiltonian of
a closed shell nucleus as
\begin{equation}
   {\mathbf H} 
   \ = \ 
   T_{k_1 k_2} \ {\mathbf a}_{k_1}^{\dag} \ {\mathbf a}_{k_2}
   \ + \ 
   \frac{1}{4} \, V_{k_1 k_2 k_3 k_4} \
   {\mathbf a}_{k_1}^{\dag} \ {\mathbf a}_{k_2}^{\dag} \
   {\mathbf a}_{k_4} \ {\mathbf a}_{k_3}
   \> ,
\end{equation}
where the Einstein summation convention is understood
and $V_{k_1 k_2 k_3 k_4}$ are fully anti-symmetric matrix elements
of the potential.
The orbits completely occupied in a non-correlated ground state
are called ``hole orbits'', those not occupied are called ``particle orbits''. 
Creation and destruction operators in these orbits are denoted
${\mathbf a}^{\dag}_p$, ${\mathbf a}^{\dag}_h$, 
and ${\mathbf a}_p$, ${\mathbf a}_h$, respectively.
Thus, we have
\begin{eqnarray}
   {\mathbf a}^{\dag}_h \ | 0 \rangle \ = \ 0
   \> ;
   \qquad
   {\mathbf a}_p \ | 0 \rangle \ = \ 0
   \> .
\end{eqnarray}

The correlated ground state is assumed to be of the form
\begin{equation}
   | \tilde 0 \rangle
   \ = \ 
   e^{{\mathbf S}^{\dag}} \ | 0 \rangle
   \ = \ 
   ( 1 \ + \ {\mathbf F}^{\dag}) \ | 0 \rangle
   \> .
\end{equation}
Here ${\mathbf S}^{\dag}$ is the operator that describes the correlation.
It is written as 
\begin{equation}
   {\mathbf S}^{\dag}
   \ = \ 
   \sum_n \ {\mathbf S}^{\dag}_n
   \> ,
\end{equation}
where 
\begin{eqnarray}
   {\mathbf S}^{\dag}_1 
   & = & 
   \sum_{ph} \ Z_{ph} \ {\mathbf a}^{\dag}_p {\mathbf a}_h 
   \> ,
   \nonumber \\
   {\mathbf S}^{\dag}_2 
   & = & 
   \frac{1}{(2!)^2} \, 
   \sum_{p_1h_1,p_2h_2} \ 
   Z_{p_1h_1,p_2h_2} \ 
   {\mathbf a}^{\dag}_{p_1} \ {\mathbf a}^{\dag}_{p_2} \
   {\mathbf a}_{h_2} \ {\mathbf a}_{h_1}  
   \> ,
   \nonumber \\
   {\mathbf S}^{\dag}_3 
   & = & 
   \frac{1}{(3!)^2} \, 
   \sum \ \cdots
   \> .
\label{eq:Sndef}
\end{eqnarray}
or correspondingly ${\mathbf F}^{\dag}$.           
The purpose of the variational calculation is to determine all the 
coefficients $Z$.

We will further assume that there exists at least one set of
wave functions, satisfying the condition that ${\mathbf S}_1$~=0.
Further down we will discuss how to construct such a set.
This set is the set of single particle mean-field wave functions.
With that basis the lowest order of correlations are the $2p2h$
correlations. Also, ${\mathbf S}_2$ and ${\mathbf F}_2$ are the same as well
as ${\mathbf S}_3$ and ${\mathbf F}_3$, whereas 
${\mathbf F}_4 = {\mathbf S}_4 + {\mathbf S}_2^2 / 2$, etc.

A variation $\delta |\tilde 0\rangle$ orthogonal to the correlated 
ground state can be constructed from any operator 
${\mathbf C^{\dag}_n}$ representing any n$p$n$h$-excitation as
\begin{equation}
   \delta \, | \tilde 0 \rangle 
   \ = \ 
   e^{- \, {\mathbf S}} \ {\mathbf C}^{\dag}_n \, e^{- \, {\mathbf S}^{\dag}} \ 
   | \tilde 0 \rangle
   \ = \ 
   e^{- \, {\mathbf S}} \ \mathbf{C}^{\dag}_n \ 
   | 0 \rangle
   \>.
\end{equation}
We have
\begin{equation}
   \langle \tilde 0 | \ \delta \, | \tilde 0 \rangle
   \ = \
   \langle 0 | \ {\mathbf C}^{\dag}_n \ | 0\rangle
   \ = \ 0
   \>.
\end{equation}
The variational principle requires that the Hamiltonian between the
ground state and such a variation vanishes. Thus, we have
\begin{equation}
   \langle \tilde 0 | \ {\mathbf H} \ \delta \, | \tilde 0 \rangle
   \ = \
   \langle 0 | \ e^{\mathbf S} \ {\mathbf H} \ e^{- \, {\mathbf S}} \ 
   {\mathbf C}^{\dag}_n \
   | 0 \rangle 
   \ = \ 0
   \>.
\label{eq:varprinc}
\end{equation}
Choosing ${\mathbf C}^{\dag}_n$ to be 1$p$1$h$, or 2$p$2$h$, ..., 
A$p$A$h$ results in a set of nonlinear coupled equations that are written out 
in more detail in the next section. 
We might say that $e^{\mathbf S} {\mathbf H}e^{- \, {\mathbf S}}$ represents the effective Hamiltonian as Eq.~(\ref{eq:varprinc}) 
represents the Hartree-Fock condition for the uncorrelated ground state 
with this effective Hamiltonian.

\subsection{The effective one-body Hamiltonian}

To simplify the accounting of number of $ph$-excitations we use 
indices for the Hamiltonian.
Explicitly we define
\begin{eqnarray}
   {\mathbf V}_{20} 
   & = & 
   \frac{1}{4} V_{p_1 p_2 h_1 h_2} 
   {\mathbf a}^{\dag}_{p_1} {\mathbf a}^{\dag}_{p_2}
   {\mathbf a}_{h_2} {\mathbf a}_{h_1} 
   \ = \ 
   {\mathbf V}^{\dag}_{02}
   \> ,
   \nonumber \\ 
   {\mathbf V}_{10} 
   & = &
   \frac{1}{2} V_{p_1 p_2 h_1 p_3}
   {\mathbf a}^{\dag}_{p_1} {\mathbf a}^{\dag}_{p_2}
   {\mathbf a}_{p_3} {\mathbf a}_{h_1} 
   \nonumber \\ &&
   \ + \ 
   \frac{1}{2} V_{p_1 h_2 h_1 h_3}
   {\mathbf a}^{\dag}_{p_1} {\mathbf a}^{\dag}_{h_2}
   {\mathbf a}_{h_3} {\mathbf a}_{h_1} 
   \ = \
   {\mathbf V}^{\dag}_{01}
   \> ,           
   \nonumber \\ 
   {\mathbf V}_{00} 
   & = &
   V_{p_1 h_2 h_1 p_2}
   {\mathbf a}^{\dag}_{p_1} {\mathbf a}^{\dag}_{h_2}
   {\mathbf a}_{p_2} {\mathbf a}_{h_1} 
   \nonumber \\ &&
   \ + \ 
   \frac{1}{4} V_{p_1 p_2 p_3 p_4}
   {\mathbf a}^{\dag}_{p_1} {\mathbf a}^{\dag}_{p_2}
   {\mathbf a}_{p_4} {\mathbf a}_{p_3} 
   \nonumber \\ &&
   \ + \
   \frac{1}{4} V_{h_1 h_2 h_3 h_4}
   {\mathbf a}^{\dag}_{h_1} {\mathbf a}^{\dag}_{h_2}
   {\mathbf a}_{h_4} {\mathbf a}_{h_3} 
   \> ,
   \nonumber \\
   {\mathbf T}_0 & = &
   T_{k_1 k_2} \
   {\mathbf a}_{k_1}^{\dag}{\mathbf a}_{k_2}
   \> .              
\end{eqnarray}
We should take note that any ${\mathbf S}_n$ commutes with ${\mathbf V}_{02}$.

We introduce the mean-field Hamiltonian ${\mathbf H}_0$ as
\begin{eqnarray}
   {\mathbf H}_0 & = &
   {\mathbf T}_0 \ + \ {\mathbf U}_0
   \>,
\end{eqnarray}
where ${\mathbf T}_0$ is the kinetic energy operator and
${\mathbf U}_0$ is the mean-field to be specified below. 
We will assume that the orbits are eigenfunctions of 
this mean-field Hamiltonian ${\mathbf H}_0$ with
\begin{equation}
   \left [ {\mathbf H}_0 , \, {\mathbf a}^{\dag}_p \right ]
   \ = \
   \epsilon_p {\mathbf a}^{\dag}_p
   \>, \qquad
   \left [ {\mathbf H}_0 , \, {\mathbf a}_h \right ]
   \ = \
   - \, \epsilon_h {\mathbf a}_h
   \> .
\label{eq:H0_eqn}
\end{equation}

We use Eq.~(\ref{eq:varprinc}) with $n=1$,  
${\mathbf C}_1^{\dag} = {\mathbf a}^{\dag}_p {\mathbf a}_h$, 
where we write 
${\mathbf a}^{\dag}_p {\mathbf a}_h \ | 0 \rangle = |1p 1h \rangle$ 
to obtain the equation establishing ${\mathbf S}_1$ 
\begin{eqnarray}
   0 & = & 
   \langle 0 | 
      \bigg \lbrace 
      {\mathbf T}_0 + 
      {\mathbf V}_{01} +
      \big[ {\mathbf S}_1, {\mathbf T}_{0} \big] +
      \big[ {\mathbf S}_1, {\mathbf V}_{00} \big] +
      \big[ {\mathbf S}_2, {\mathbf V}_{10} \big] +
   \nonumber \\ &&
      \big[ {\mathbf S}_3, {\mathbf V}_{20} \big]
     + \frac{1}{2} 
       \Big[ {\mathbf S}_1, \big[ {\mathbf S}_1, {\mathbf V}_{10} \big] \Big] 
     + \Big[ {\mathbf S}_1, \big[ {\mathbf S}_2, {\mathbf V}_{20} \big] \Big] 
   \nonumber \\ &&
     + \frac{1}{6} 
       \bigg[ {\mathbf S}_1, 
              \Big[ {\mathbf S}_1, 
                    \big[{\mathbf S}_1, {\mathbf V}_{20}\big] 
              \Big]
       \bigg] 
     \bigg \rbrace 
   | 1p1h \rangle
   \> .
   \label{eq:1p1h}
\end{eqnarray}
There are similar equations that determine 
${\mathbf S}_2, {\mathbf S}_3, \ldots$.
While these equations hold in any basis, there is one basis of
particular convenience. 
This is the \emph{maximum overlap basis} in which ${\mathbf S}_1$ vanishes.
Equation~(\ref{eq:1p1h}) results in the solution ${\mathbf S}_1=0$ if the terms
that do not contain ${\mathbf S}_1$ vanish. 
The mean-field basis is determined by the condition of the vanishing 
of ${\mathbf S}_1$ and in the mean-field basis we must have
\begin{eqnarray}
   0 & = \
   \langle 0 | \ 
       \bigg \lbrace 
     &
             ( {\mathbf H}_0 - {\mathbf U}_0 ) 
             \ + \ {\mathbf V}_{01}
   \nonumber \\ &&
             \ + \ \big[ {\mathbf S}_2, {\mathbf V}_{10} \big] 
             \ + \ \big[ {\mathbf S}_3, {\mathbf V}_{20} \big] \ 
       \bigg \rbrace \ 
   | 1p1h \rangle
   \> .
\label{eq:S1eqn_var1}
\end{eqnarray}
Using Eq.~(\ref{eq:H0_eqn}), we can show that the expectation value
$\langle 0 | {\mathbf H}_0 | 1p1h \rangle$ vanishes. 
Therefore, Eq.~(\ref{eq:S1eqn_var1}) becomes 
\begin{equation}
   \langle 0 | {\mathbf U}_0 | 1p1h \rangle
   = 
   \langle 0 |
       \bigg \lbrace
               {\mathbf V}_{01}
             + \big[ {\mathbf S}_2, {\mathbf V}_{10} \big]
             + \big[ {\mathbf S}_3, {\mathbf V}_{20} \big]
       \bigg \rbrace
   | 1p1h \rangle
   \> .
\label{eq:S1eqn}
\end{equation}

Thus those terms establish the elements in the one-body Hamiltonian
matrix that connect $p$ and $h$ orbits. 
The equations establishing the higher order correlations in the mean-field
basis are
\begin{eqnarray}
   0 & = & 
   \langle 0 | 
       \bigg \lbrace {\mathbf V}_{02}
                     + \big[ {\mathbf S}_2, {\mathbf V}_{00} \big] 
                     + \big[ {\mathbf S}_2, {\mathbf H}_{0} \big] 
                     - \big[ {\mathbf S}_2, {\mathbf U}_{0} \big] 
   \nonumber \\ &&
                     + \big[ {\mathbf S}_3, {\mathbf V}_{10} \big]
                     + \big[ {\mathbf S}_4, {\mathbf V}_{20} \big]
                     + \frac{1}{2} \Big[ {\mathbf S}_2, 
                                         \big[ {\mathbf S}_2, {\mathbf V}_{20} \big] 
                                   \Big] 
       \bigg \rbrace 
   | 2p2h \rangle
   \> ,
   \nonumber \\ &&
   \\
   0 & = & 
   \langle 0| 
       \bigg \lbrace \big[ {\mathbf S}_2, {\mathbf V}_{01} \big]
                     + \big[ {\mathbf S}_3, {\mathbf V}_{00} \big]
                     + \big[ {\mathbf S}_3, {\mathbf H}_{0} \big]
                     - \big[ {\mathbf S}_3, {\mathbf U}_{0} \big]
   \nonumber \\ &&
                     + \big[ {\mathbf S}_4, {\mathbf V}_{10} \big] 
                     + \big[ {\mathbf S}_5, {\mathbf V}_{20} \big]
                     + \frac{1}{2} \Big[ {\mathbf S}_2,
                                         \big[ {\mathbf S}_2, {\mathbf V}_{10} \big] 
                                   \Big] 
   \nonumber \\ &&
                     + \Big[ {\mathbf S}_3,
                             \big[{\mathbf S}_2, {\mathbf V}_{20} \big] 
                       \Big]
       \bigg \rbrace 
   | 3p3h \rangle
   \> ,
   \\
   0 & = & 
   \langle 0| 
       \bigg \lbrace \big[ {\mathbf S}_3, {\mathbf V}_{01} \big]
                     + \big[ {\mathbf S}_4, {\mathbf V}_{00} \big]
                     + \big[ {\mathbf S}_4, {\mathbf H}_{0} \big]
                     - \big[ {\mathbf S}_4, {\mathbf U}_{0} \big]
   \nonumber \\ &&
                     + \big[ {\mathbf S}_5, {\mathbf V}_{10} \big]
                     + \big[ {\mathbf S}_6, {\mathbf V}_{20} \big]
                     + \frac{1}{2} \Big[ {\mathbf S}_2, 
                                         \big[{\mathbf S}_2, {\mathbf V}_{00} \big]
                                   \Big] 
   \nonumber \\ &&
                     + \Big[ {\mathbf S}_3, 
                             \big[ {\mathbf S}_2, {\mathbf V}_{10} \big]
                       \Big]
                     + \Big[ {\mathbf S}_4,
                             \big[{\mathbf S}_2, {\mathbf V}_{20} \big] 
                       \Big]
   \nonumber \\ &&
                     + \frac{1}{2} \Big[ {\mathbf S}_3,
                                         \big[ {\mathbf S}_3, {\mathbf V}_{20} \big] 
                                   \Big] 
                     + \frac{1}{6} \bigg[ {\mathbf S}_2, 
                                          \Big[ {\mathbf S}_2,
                                                \big[{\mathbf S}_2, {\mathbf V}_{20} \big]
                                          \Big]
                                   \bigg]
       \bigg \rbrace 
   | 4p4h \rangle
   \> .
   \nonumber \\
\label{eq:S4_eqn}
\end{eqnarray}
At this point we will assume that the orbits are eigenfunctions to the
single particle Hamiltonian ${\mathbf H}_0$. This allows us to solve these
equations as
\begin{eqnarray}
   \langle 0| {\mathbf S}_2 | 2p2h \rangle & = & 
      - \langle 0 | 
            \bigg \lbrace {\mathbf V}_{02}
                          + \big[ {\mathbf S}_2, {\mathbf V}_{00} \big] 
                          - \big[ {\mathbf S}_2, {\mathbf U}_{0} \big] 
   \nonumber \\ &&
                          + \big[ {\mathbf S}_3, {\mathbf V}_{10} \big]
                          + \big[ {\mathbf S}_4, {\mathbf V}_{20} \big]
   \nonumber \\ &&
                          +\frac{1}{2} \Big[ {\mathbf S}_2,
                                             \big[ {\mathbf S}_2, {\mathbf V}_{20} \big]
                                       \Big] 
            \bigg \rbrace \frac{1}{{\mathbf H}_0}
        | 2p2h \rangle
   \> ,
\label{eq:2p2h}
\end{eqnarray}
\begin{eqnarray}
   \langle 0| {\mathbf S}_3 | 3p3h \rangle & = & 
      - \langle 0 |
            \bigg \lbrace \big[ {\mathbf S}_2, {\mathbf V}_{01} \big]
                          + \big[ {\mathbf S}_3, {\mathbf V}_{00} \big]
                          - \big[ {\mathbf S}_3, {\mathbf U}_{0} \big]
   \nonumber \\ &&
                          + \big[ {\mathbf S}_4, {\mathbf V}_{10} \big] 
                          + \big[ {\mathbf S}_5, {\mathbf V}_{20} \big] 
                          + \frac{1}{2} \Big[ {\mathbf S}_2,
                                              \big[{\mathbf S}_2, {\mathbf V}_{10} \big] 
                                        \Big] 
   \nonumber \\ &&
                          + \Big[ {\mathbf S}_3,
                                  \big[{\mathbf S}_2, {\mathbf V}_{20} \big] 
                            \Big]
            \bigg \rbrace \frac{1}{{\mathbf H}_0}
        | 3p3h \rangle 
   \> .
\label{eq:3p3h}
\end{eqnarray}
A similar equation allows us to isolate ${\mathbf S}_4$ using Eq.~(\ref{eq:S4_eqn}).

If we restrict ourselves to a situation in which only $2p2h$ correlations 
are present, ${\mathbf S}_2$ and ${\mathbf F}_2$ become identical. 
Then, Eq.~(\ref{eq:2p2h}) can be written in its adjoint form as
\begin{equation}
   {\mathbf F}^{\dag}_2 
   \ = \ 
   - \ \frac{1}{{\mathbf H}_0} \ ( {\mathbf V} 
                               \, + \, {\mathbf V} \, {\mathbf F}^{\dag}_2 )
   \> .
\label{eq:2p2hadj}
\end{equation}
This equation is essentially the Bethe-Goldstone equation for the
${\mathbf G}$-matrix which is normally written as~\cite{ref:Bishop}
\begin{equation}
   {\mathbf G}^{\dag}(\omega)
   \ = \
   {\mathbf V} \ \Bigl[ {\mathbf 1}
                        + \frac{1}{\omega -{\mathbf H}_0} \, {\mathbf G}^{\dag}(\omega)
                 \Bigr]
   \> .
\end{equation}
Multiplying this equation from the left with
$1/(\omega -{\mathbf H}_0)$,
using
${\mathbf F}^{\dag}_2 = \left [ \omega -{\mathbf H}_0) \right ]^{-1} 
                                {\mathbf G}^{\dag}(\omega)$ 
and setting $\omega=0$, 
results in Eq.~(\ref{eq:2p2hadj}) for ${\mathbf F}^{\dag}_2$.
Thus, one of the essential parts of the coupled cluster expansion is that
we calculate the ground state two-body $G$-matrix inside the nucleus where 
it is to be applied. 
The additional terms in our coupled equations indicate the need for appropriate
corrections in the $G$-matrix for the presence of $3p3h$, $4p4h$, $etc.$
correlations. 
In most cases the ${\mathbf G}$-matrix is calculated in nuclear matter
and then applied to finite nuclei. However, in nuclear matter the
correlation function extends up to infinity and thus cannot be applied
to a finite nucleus.

We estimate that in our basis there are about $10^6$ $2p2h$
configurations and about $10^{10}$ $3p3h$ configurations.
While the number of $2p2h$ configurations is quite accessible, the
number of $3p3h$ configurations is prohibitively large, and we cannot
store all these numbers. Thus we have to implicitly correct for the
presence of these correlations. 
We do this by inserting the solutions for ${\mathbf S}_n$ with n$\ge$3 
back into the equations and thereby obtaining a perturbation expansion
in $1/E_{ph}$. We write this equation out up to second order for the
Eq.~(\ref{eq:S1eqn}) establishing the mean-field 
\begin{eqnarray}
   &&
   \langle 0| {\mathbf U}_0 | 1p1h \rangle
   \ = \
   \langle 0| {\mathbf V}_{01} | 1p1h \rangle 
   + \langle 0| \big[ {\mathbf S}_2, {\mathbf V}_{10} \big]| 1p1h \rangle 
   \nonumber \\ &&
   - \langle 0| \Big[ \big[ {\mathbf S}_2, {\mathbf V}_{01}\big],
                            \frac{1}{{\mathbf H}_0}{\mathbf V}_{20}
                \Big]
     | 1p1h \rangle 
   \nonumber \\ &&
   - \frac{1}{2} \langle 0|
                     \bigg[ \Big[ {\mathbf S}_2, 
                                  \big[ {\mathbf S}_2, {\mathbf V}_{10} \big]
                            \Big], 
                            \frac{1}{{\mathbf H}_0} {\mathbf V}_{20}
                     \bigg]
                 | 1p1h \rangle 
   \nonumber \\ &&
   + \langle 0|
         \bigg[ \Big [ \big[ {\mathbf S}_2, {\mathbf V}_{01} \big],
                       \frac{1}{{\mathbf H}_0} {\mathbf V}_{00} 
                \Big],
                \frac{1}{{\mathbf H}_0} {\mathbf V}_{20}
         \bigg]
     | 1p1h \rangle 
   \> .
   \label{eq:S1eqn2}
\end{eqnarray}

This equation establishes the matrix elements of the single particle
Hamiltonian ${\mathbf H}_0$ between particle and hole
orbits. The matrix elements between hole and hole orbits or between
particle and particle orbits are not defined, and any definition may
be chosen. As long as ${\mathbf U}_0$ is explicitly kept on the right
hand side of Eqs.~(\ref{eq:2p2h}, \ref{eq:2p2hadj}) the explicit choice 
is merely a question of how fast the resulting series will converge.
However, a reasonable choice appears to be that form that we
obtain if we replace in the matrix elements obtained in Eq.~(\ref{eq:S1eqn2}) 
the hole orbit with a particle orbit in order to get the matrix elements between
particle and particle orbits and we change the particle orbit into
a hole orbit in order to get the matrix elements between hole and hole orbits. 
Reference~\cite{ref:terms} gives a detailed account of the contributions 
included in our mean-field as given by Eq.~(\ref{eq:S1eqn2}). 
Our choice for the other matrix elements corresponds
simply in turning the hole line into a particle line or vice versa.

The mean-field orbits are the eigenvectors of this matrix,
and the eigenvalues are the single particle energies.
This procedure now fully defines the mean-field used here even though
its definition is not unique.

To obtain values for the amplitudes of ${\mathbf S}_2$ we start with
Eq.~(\ref{eq:2p2h}) and replace ${\mathbf S}_3$ by Eq.~(\ref{eq:3p3h}) 
and correspondingly ${\mathbf S}_4$. 
We continue with these replacements, resulting 
again in an expansion in $1/E_{ph}$ for ${\mathbf S}_2$.

The calculation of the correlations breaks down into two steps:
In step one the two-body correlations are computed according to 
Eq.~(\ref{eq:2p2h}) 
where the curly brackets give  the various contributions to the 
effective $\langle p_1h_1|V_{ph}|h_2p_2 \rangle$ matrix element. In step two
the single particle energy tensor is calculated from the
kinetic energy, the direct potential energy, and the correlation
energy. The eigenvalues and the eigenvectors of the energy tensor
give the new single particle energies and the new single particle
functions. The two steps are iterated until a self-consistent solution
is obtained, i.e. the energy tensor is diagonal.

As we use an infinite expansion we must truncate that expansion.
We find that terms connecting to effective one-body terms give 
the major contributions, two-body terms are less important. 
Generally, we have included all terms that are written as product
of three two-body operators. 
Of those terms written as product of four two-body operators we have included 
those terms where at least two operators connect to a one-body operator. 
This results in all the quenching terms for the products of two two-body
operators. 
We have also included those terms that look like the terms of the products 
of three two-body operators with a renormalized interaction.
We found that these corrections
are equally significant for those terms arising from ${\bf S}_4$
and those arising from ${\bf S}_3$. 
Left out are mostly those terms that required several angular momentum 
recouplings. 
Contributions from ${\bf S}_5$ show up only if one considers products of 
more than four two-body operators.
In the Ref.~\cite{ref:terms} we have listed explicitly
the relations and approximations that we have used in this calculation.

\section{Numerical calculations for $^{16}$O}

We have solved the main Eq.~(\ref{eq:S1eqn2}) that determines the 
$2p2h$-amplitudes
and thus essentially the ground state $G$-matrix for $^{16}$O
in a space of 50 $\hbar\omega$ with a harmonic oscillator length
parameter b=0.8 fm, excluding those orbits with $l \ge 13$. 
Corrections for $3p3h$ correlations were included in a reduced space of 
30 $\hbar\omega$ and $l \le 6$, 
while correlations due to $4p4h$ correlations were included in the full space.
We used the Argonne $v_{18}$ potential~\cite{ref:WiringaETal} 
to generate the matrix elements~\cite{ref:2b-me}. 
For comparison, results for the Argonne $v_8$ and $v_{14}$ potentials are reported.
Here, the Argonne $v_8$ potential is the reprojection of the Argonne $v_{14}$ potential 
in the sense of reference~\cite{ref:Pudlineretal_2}.

The Hamiltonian is given in the center-of-mass as
\begin{eqnarray}
   H_{int} 
   & = &
   \sum_i^A 
   \frac{1}{2m} \vec {\mathbf p}^2_i
   \, + \,
   \sum_{i<j}^A 
   V(\vec {\mathbf r}_i - \vec {\mathbf r}_j )
   \, - \, T_{CM}
   \> ,
\end{eqnarray}
where $T_{CM} \equiv \vec P^2/2M$ is the kinetic energy operator of the 
center-of-mass, and $M$ is the total mass.
This represents the energy in the center-of-mass frame. 
It can be rewritten as
\begin{eqnarray} 
   H_{int} 
   & = &
   \left ( 1 - \frac{1}{A} \right ) \ 
   \sum_{i=1}^A \ 
   \frac{1}{2 m} \ p_i^2
   \nonumber \\ &&
   \ + \ 
   \sum_{i<j=1}^A \ 
   \left [ 
      V(\vec {\mathbf r}_i \, - \, \vec {\mathbf r}_j )
      \ - \ \frac{\vec {\mathbf p}_i \cdot \vec {\mathbf p}_j}{Am} \ 
   \right ]
   \> .
\end{eqnarray}
The last term is treated as part of the internal potential, and the 
antisymmetric matrix
elements of the internal potential include this center-of-mass correction term.

\subsection{Ground state expectation values for arbitrary operators}

Ground state expectation values can be evaluated by introducing the
operator ${\mathbf \tilde S}^{\dag}$, as described e.g. in the review
by Bishop \cite{ref:Bishop}. We take the ground state as
\begin{equation}
   | \tilde 0\rangle
   \ = \
   e^{{\mathbf S}^{\dag}} \ | 0\rangle
   \> .
\end{equation}
We have defined $\lbrace{\mathbf C}^{\dag}_n\rbrace$ as the complete set of
1$p$1$h$, 2$p$2$h$, ..., A$p$A$h$ excitations.
The normalized expectation value  $\bar a$
of any operator ${\mathbf A}$, $\bar a = \langle {\mathbf A} \rangle$, 
can be worked out as
\begin{equation}
   \bar a 
   \ = \
   \frac{ \langle 0 | \ e^{\mathbf S} \ {\mathbf A} \ e^{ {\mathbf S}^{\dag} } \ 
          | 0 \rangle }
        { \langle \tilde 0 \, | \, \tilde 0 \rangle}  
   \ = \ 
   \frac{ \langle 0 | \ e^{\mathbf S} \ {\mathbf A} \ e^{- \, {\mathbf S}} \ e^{\mathbf S} \ 
                        e^{ {\mathbf S}^{\dag} } \ 
          | 0 \rangle }
        { \langle \tilde 0 \, | \, \tilde 0 \rangle}  
   \> .
\end{equation}
By inserting the unity operator in the 
$\{ {\mathbf C}^{\dag}_n \ | 0 \rangle \}$ basis, we obtain
\begin{eqnarray}
   &&
   \bar a 
   \ = \
   \langle 0 | \ e^{\mathbf S} \ {\mathbf A} \ e^{- \, {\mathbf S}} \ | 0 \rangle
   \nonumber \\ &&
   \ + \ 
   \sum_n \ 
   \langle 0 | \ e^{\mathbf S} \ {\mathbf A} \ e^{- \, {\mathbf S}} \ {\mathbf C}^{\dag}_n \ 
   | 0 \rangle \ 
   \frac{ \langle 0 | \ e^{\mathbf S} \ {\mathbf C}_n \ e^{ {\mathbf S}^{\dag} } \ 
          | 0 \rangle }
        { \langle \tilde 0 \, | \, \tilde 0 \rangle}  
   \> .
\end{eqnarray}
The expectation value on the right is by definition $\bar c_n$, the 
expectation value of ${\mathbf C}_n$. 
Thus we can define the new operator
\begin{equation}
   \tilde {\mathbf S}^{\dag}
   \ = \ 
   \sum_{n=1}^\infty \ \bar c_n \ {\mathbf C}^{\dag}_n
   \> .
\label{eq:Stil}
\end{equation}
With this, the expectation value for any operator can be expressed as
\begin{equation}
   \bar a \ = \ 
   \langle 0 | \ e^{\mathbf S} \ {\mathbf A} \ e^{- \, {\mathbf S}} \ 
   \left ( 1 \, + \, \tilde {\mathbf S}^{\dag} \right )
   | 0 \rangle
   \> .
\label{eq:Abar}
\end{equation}
The operators ${\mathbf \tilde S}^{\dag}$ can be obtained by solving
Eq.~(\ref{eq:Stil}) in an iterative fashion. Explicitly we write
${\mathbf \tilde S}^{\dag}$ in the same form as Eq.~(\ref{eq:Sndef}), but
use the amplitudes $S$ instead of $Z$ to distinguish it from
${\mathbf S}^{\dag}$.

\subsection{Ground state binding energy}

We first apply the above formalism  to the ground state binding energy. 
The expectation value of the internal Hamiltonian can be written as
\begin{equation}
   \bar e 
   \ = \ 
   \langle 0 | \
   e^{\mathbf S} \ {\mathbf H}_{int} \ e^{- \, {\mathbf S}} \ 
   | 0 \rangle
   \>.
\end{equation}
Because of the Hartree-Fock (HF) condition expressed 
in Eq.~(\ref{eq:varprinc}) the terms
involving ${\mathbf \tilde S}^{\dag}$ vanish and we get
\begin{equation}
   \bar e 
   \ = \ 
   \langle 0 | \
   e^{\mathbf S} \ {\mathbf H} \ e^{- \, {\mathbf S}} \ 
   | 0 \rangle
   \>.
\end{equation}
Assuming that ${\mathbf H}$ is at most a two-body operator and taking into
account that ${\mathbf S}_1$ vanishes, we write this as
\begin{equation}
   \langle E \rangle
   \ = \ 
   \bar e \ = \ 
   \langle  0 | \ {\mathbf H} \ |0 \rangle 
   \ + \ 
   \langle 0 | \ {\mathbf S}_2 \ {\mathbf V}_{20} \ | 0 \rangle
   \>.
\end{equation}
When we evaluate the expectation values of the operators in the above equation, 
we have to consider that the hole orbits are
not diagonal with respect to any of these operators. 
Also, as we have used the HF condition this expression does not give an upper 
limit of the ground state energy unless we are exactly at the minimum.
In terms of matrix elements the energy can be written as
\begin{eqnarray}
   \langle E \rangle 
   & = &
   \sum_{h_1 h_2} \ T_{h_1,h_2}
   \ + \ \frac{1}{2} \ \sum_{h_1 h_2} \ V_{h_1 h_2, h_1 h_2}
   \nonumber \\ &&
   \ + \ \frac{1}{4} \ \sum_{p_1 p_2 h_1 h_2} \ Z_{p_1 p_2 , h_1 h_2} \ V_{p_1 p_2 , h_1 h_2}
   \>.
\label{eq:binding}
\end{eqnarray}
The use of Eq.~(\ref{eq:binding}) for the energy implies that all
the correlations are present and satisfy the Eqs.~(\ref{eq:varprinc}).
In that case the resulting energy could be taken
as an upper bound to the true energy. However, in solving these equations
we have ignored some of the couplings back into ${\bf S}_2$.
As a result Eq.(\ref{eq:binding}) is no longer exact and the feature
of being an upper bound to the true energy is lost.
We believe that the errors due to the truncations are reasonably small.
However, more experience is needed with some of the terms left out
in order to reduce the uncertainties in the result.
Table~\ref{tab:erms_occ_2b} shows the resulting binding energy 
for the Argonne $v_8$, $v_{14}$ and $v_{18}$ potentials.

\begin{figure}
   \epsfxsize = 3.0in
   \centerline{\epsfbox{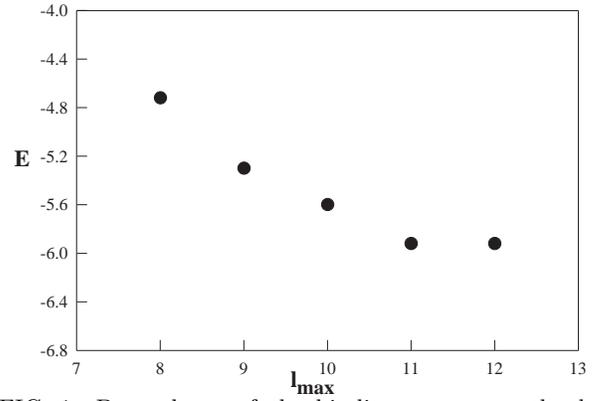}}
   \caption{Dependence of the binding energy on the $l_{max}$ cut-off
            for $n_{max}$=25.}
   \label{fig:l_depv18}
\end{figure}

\begin{figure}
   \epsfxsize = 3.0in
   \centerline{\epsfbox{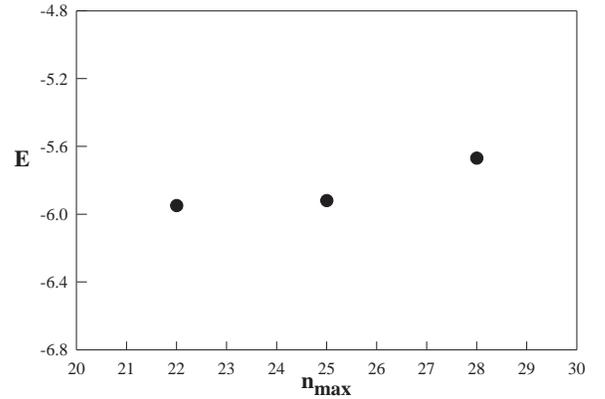}}
   \caption{Dependence of the binding energy on the $n_{max}$ cut-off
            for $l_{max}$=11.}
   \label{fig:n_depv18}
\end{figure}

Our Hilbert space is controlled by two cut-off parameters $l_{max}$ and
$n_{max}$, which are related to the size of the Hilbert space as 
$N_{max}=2n_{max}+l_{max}$. 
We have studied the dependence of the binding energy on these two parameters. 
Figure~\ref{fig:l_depv18} displays the $l_{max}$ dependence and shows 
reasonable convergence for $l_{max}=11$. 
This dependence was mapped out with $n_{max}$=25.
We have also checked the sensitivity of the binding energy 
with respect to $n_{max}$ with a fixed $l_{max}$. 
In this case, we find that for our choice of a 30 $\hbar \omega$ $3p3h$ space, 
the binding energy is independent of $n_{max}$ for an $n_max$ value 
between 22 and 25. 
As we go to larger $n_{max}$ values however, 
we start seeing the effect of having a smaller $3p3h$ space 
and the approximation starts breaking down.
The dependence of the binding energy on the $n_{max}$ cut-off is depicted in
Fig.~\ref{fig:n_depv18} for $l_{max}$=11.

\subsection{Ground state one-body density}

Next, we apply this procedure to the ground state one-body density. The
expressions necessarily look more complex as we cannot apply the
HF condition to simplify the expressions. 

By definition, the ground state one-body density is introduced as
\begin{equation}
   \rho(\vec r) 
   \ = \ 
   \sum_{k=1}^A \ 
   \langle \ \delta(\vec r - \vec r_k) \ 
   \rangle 
   \> .
\label{oneden_def}   
\end{equation}
Since we are dealing with a spherically symmetric system, we shall integrate 
out the angular degrees of freedom of the system.
Then, we write the density operator in the second quantisation representation 
as
\begin{equation}
   \rho^{op}(r) \ = \
   \sum_{\alpha \beta} \ 
   \rho_{\alpha \beta}(r) \ 
   {\mathbf a}^{\dag}_\alpha \, {\mathbf a}_\beta
   \> .
\end{equation}
Here we use $\rho_{\alpha \beta}(r) = R_\alpha(r) \, R_\beta(r)$ 
to denote the radial part of the expectation value 
$\langle \alpha \, | \ \delta(\vec r - \vec r') \ | \, \beta \rangle$.
Thus, we can write the ground state density as
\begin{equation}
   \rho(r) 
   \ = \ 
   \sum_{\alpha \beta} \ d_{\alpha \beta} 
                       \ R_\alpha(r) \ R_\beta(r)
   \>,
\end{equation}
where 
\begin{equation}
   d_{a,b}
   \ = \
   \langle \ {\mathbf a}^{\dag}_\alpha \ {\mathbf a}_\beta \ 
   \rangle
\label{eq:denmat}
\end{equation}
is the density matrix. 
The density matrix is a real symmetric matrix with positive definite 
eigenvalues. We can make a basis transformation such that
the density matrix becomes diagonal. This basis represents the
``natural'' orbits. In this basis the density becomes
\begin{equation}
   \rho(r) \ = \ 
   \sum_a \ v^{nat}_a \ \left [ R^{nat}_a(r) \right ]^2
   \> .
\end{equation}
Here $v^{nat}_a$ represents the occupation probability of these
natural orbits. This is the only basis in which occupation probabilities
have a meaning.

To calculate the one-body density matrix we use Eq.~(\ref{eq:denmat}) 
with Eq.~(\ref{eq:Abar})
\begin{eqnarray}
   &&
   d_{\alpha \beta} 
   \ = \ 
   \langle 0 \, | \ 
       {\mathbf a}^{\dag}_\alpha {\mathbf a}_\beta
   \ | \, 0 \rangle
   \ + \ 
   \langle 0 \, | \
       \left [ {\mathbf S}_2, \ {\mathbf a}^{\dag}_\alpha {\mathbf a}_\beta
       \right ] \ 
       \tilde {\mathbf S}_2^{\dag}
   \ | \, 0 \rangle
   \nonumber \\ &&
   \ + \ 
   \langle 0 \, | \ 
       {\mathbf a}^{\dag}_\alpha {\mathbf a}_\beta \ 
       \tilde {\mathbf S}_1^{\dag}
   \ | \, 0 \rangle
   \ + \ 
   \langle 0 \, | \
       \left [ {\mathbf S}_2, \ {\mathbf a}^{\dag}_\alpha {\mathbf a}_\beta 
       \right ] \ 
       \tilde {\mathbf S}_1^{\dag}
   \ | \, 0 \rangle
   \nonumber \\ &&
   \ + \ 
   \langle 0 \, | \
       \left [ {\mathbf S}_3, \ {\mathbf a}^{\dag}_\alpha {\mathbf a}_\beta 
       \right ] \ 
       \tilde {\mathbf S}_2^{\dag}
   \ | \, 0 \rangle
   \ + \ 
   \langle 0 \, | \
       \left [ {\mathbf S}_3, \ {\mathbf a}^{\dag}_\alpha {\mathbf a}_\beta 
       \right ] \ 
       \tilde {\mathbf S}_3^{\dag}
   \ | \, 0 \rangle
   \nonumber \\ &&
   \ + \ 
   \frac{1}{2} \ 
   \langle 0 \, | \ 
       \left [ {\mathbf S}_2, \ 
               \left [ {\mathbf S}_2, \ {\mathbf a}^{\dag}_\alpha {\mathbf a}_\beta \right ] 
       \right ] \ 
       \tilde {\mathbf S}_3^{\dag} 
   \ | \, 0 \rangle
   \ + \ 
   \cdots
   \>.
\end{eqnarray}
In evaluating terms of equal order we have to consider ${\mathbf S}_2$ to be
of order $1/E_{ph}$, ${\mathbf S}_3$ to be of order $1/E_{ph}^2$, ${\mathbf
S}_4$ to be of order  $1/E_{ph}^3$, etc. To account for ${\mathbf S}_3$ and
${\mathbf S}_4$ we have to use the expansions of Eq.~(\ref{eq:2p2h}). As 
${\mathbf S}_1$ vanishes, ${\mathbf \tilde S}_1$ is of the same order as
${\mathbf S}_3$. Otherwise ${\mathbf \tilde S}_n$ is of the same order as 
${\mathbf S}_n$.

The one-body density calculated this way is not the density in the
center-of-mass frame of the nucleus as the wave function represents a nucleus that
has a residual center-of-mass motion. 
The form factor measured in experiments is the form
factor in the center-of-mass frame which we can write as
\begin{equation}
   f(q) \ = \ 
   \sum_j \ 
   \langle \ e^{i \vec q \cdot (\vec r_j - \vec R_{cm})} \ \rangle
   \ = \
   \langle \ e^{i  \vec q \cdot \vec r_j}    \ \rangle \ 
   \langle \ e^{-i \vec q \cdot \vec R_{cm}} \ \rangle
   \> .
   \label{eq:formfact}
\end{equation}
i.e. the form factor calculated from the one-body density factorizes
into the form factor in the center-of-mass frame and the form factor of the
center-of-mass motion
$F_{cm}(q)=\langle e^{-i\vec q \cdot \vec R_{cm}} \rangle$.

It has been shown that for a pure uncorrelated harmonic oscillator 
wave function this form factor can be written as
$F_{cm}(q)=exp(-b^2q^2/4A)$
where $b$ is the harmonic oscillator length parameter. 
There are two reasons why this does not apply here. 
First, we have self-consistent mean-field wave
functions and not harmonic oscillator wave functions and second,
we have not a single Slater determinant but a sum over many due to
the correlations. For these reasons we chose to evaluate the
form factor directly using Eq.~(\ref{eq:formfact}). 
By writing $\vec R_{cm}= {1\over{A}}\sum_i\vec r_i$, 
the Eq.~(\ref{eq:formfact}) is expanded
into n-body terms. This expansion was checked for the case of
a single Slater determinant of harmonic oscillator functions against
the exact result. It was found that this expansion is quite
satisfactory if all terms up to three-body terms are kept~\cite{ref:CM_corr}.

\begin{figure}
   \epsfxsize = 3.0in
   \centerline{\epsfbox{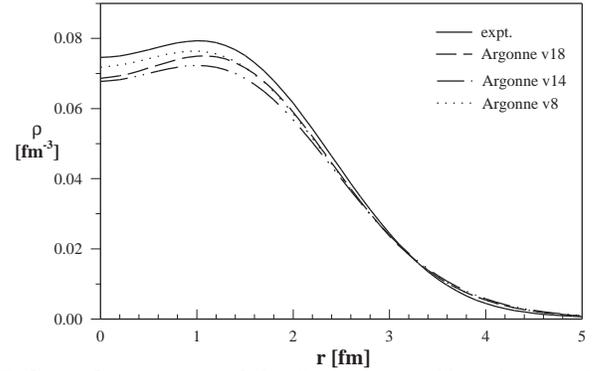}}
   \caption{Comparison of the ``experimental" 
            with the calculated charge density, 
            for the Argonne $v_8$, $v_{14}$ and $v_{18}$ potentials, respectively.}
   \label{fig:oneden_2body}
\end{figure}

Fig.~\ref{fig:oneden_2body} shows the
calculated charge density after folding the proton point density
with the charge density of the proton and folding the neutron
point density with the charge density of the neutron. 
As this expansion is accurate up to terms of order $q^4$, 
it encompasses the result of the rms-radius. 
At the present time no corrections due to the meson-exchange 
charge density are taken into account.
The resulting charge radii are shown in Table~\ref{tab:erms_occ_2b}
for the Argonne $v_8$, $v_{14}$ and $v_{18}$ potentials, respectively,
and are reasonably close to the experimental one. 

In the calculation of the natural orbits we also generate the occupation
probabilities for the orbits above the Fermi level. 
For the Argonne $v_8$, $v_{14}$ and $v_{18}$ potentials the occupation probabilities
of the 1$d_{5/2}$ and the 2$s_{1/2}$ proton orbits are summarized in 
Table~\ref{tab:erms_occ_2b} and appear to be consistent
with the experiment~\cite{ref:LeuschnerETal}, 
which establishes a lower limit for these values.

\subsection{Ground state two-body density}

A direct presentation of the short range correlation can be seen
in the ground state two-body density.
We start with the ground state two-body density definition 
\begin{equation}
   \rho(\vec r_1, \vec r_2)
   \ = \
   \sum_{i j} \ 
   \langle
       \tilde 0 \, | \ \delta(\vec r_1 - \vec r_i) \ 
                       \delta(\vec r_2 - \vec r_j) \
       | \, \tilde 0
   \rangle
   \> .
\label{eq:twoden_def}
\end{equation}
In the second quantization representation the two-body density operator 
can be written as
\begin{eqnarray}
   \rho^{op}(\vec r_1, \vec r_2)
   & = & 
   \sum_{\alpha \beta \gamma \delta} \ 
   \langle \alpha \beta \, | \, \rho(\vec r_1, \vec r_2) \, | \, \gamma \delta \rangle \ 
   {\mathbf a}^{\dag}_\alpha \ {\mathbf a}^{\dag}_\beta \ 
   {\mathbf a}_\delta \ {\mathbf a}_\gamma
   \>.
   \nonumber \\
\end{eqnarray}
Using the completeness relationship of the spherical harmonics
we can evaluate the matrix element $\rho_{\alpha \beta \gamma \delta}$
\begin{eqnarray} 
   &&
   \langle \alpha \beta \, | \, \rho(\vec r_1, \vec r_2) \, | \, \gamma \delta \rangle \ 
   \nonumber \\ &&
   = \ 
   \sum_{l_1 m_1} \,
      R_\alpha(r_1) \, R_\gamma(r_1) \, Y_{l_1 m_1}^*(\hat r_1) \, 
      \langle j_\alpha m_\alpha \, | \, Y_{l_2 m_2}       \, | \, j_\gamma m_\gamma \rangle 
   \nonumber \\ && 
   \times \
   \sum_{l_2 m_2} \,
      R_\beta(r_2) \, R_\delta(r_2) \, Y_{l_2 m_2}(\hat r_2) \, 
      \langle j_\beta  m_\beta  \, | \, Y_{l_2 m_2}^* \, | \, j_\delta m_\delta \rangle 
   \>.
   \nonumber \\
\end{eqnarray}
In order to be consistent with the phase convention of the 
two-body potential matrix elements, 
we couple the two-body density matrix elements using the 
{\emph ph} angular momentum coupling conventions~\cite{ref:2b-me}.
Then, the angular momentum coupled density is
\begin{eqnarray}
   &&
   \langle 
      (\alpha \bar \gamma)_\lambda \, | \, \rho^{\lambda \mu}(\vec r_1, \vec r_2) 
                                   \, | \, (\delta \bar \beta)_\lambda
   \rangle 
   \nonumber \\ &&
   = \
     \rho^{\lambda}_{\alpha \gamma}(r_1) \ 
     \rho^{\lambda}_{\delta \beta}(r_2) \ 
     \frac{1}{2 \lambda + 1} \ 
     Y_{\lambda \mu}^{*}(\hat r_1) \ Y_{\lambda \mu}(\hat r_2)
   \>.
\end{eqnarray}
Here we have introduced the one-body multipole density 
$\rho^{\lambda}_{\alpha \beta}(r)$ which is
\begin{eqnarray}
   & = &
   (-)^{j_\alpha + 1/2} \ R_\alpha(r) \, R_\beta(r) \ 
   \langle j_\alpha \, \| \, Y_{\lambda\mu} \, \| \, j_\beta \rangle
   \nonumber \\ 
   & = &
   (-)^{\lambda+1} \ 
   \sqrt{(2 j_\alpha + 1) \, (2 j_\beta + 1)} \
   \nonumber \\       && 
   \langle j_\alpha 1/2 \, j_\beta -1/2 \ | \ \lambda 0 \rangle \ 
   R_\alpha(r) \, R_\beta(r)
\end{eqnarray}
if $l_\alpha + l_\beta + \lambda$ is even, and zero otherwise.

For a spherically symmetric (spin=0) nucleus it is more relevant to 
calculate $\rho(r_1,r_2,\theta_{12})$ as due to the spherical symmetry
the two-body density is dependent on the direction of $\vec r_1$ 
relative to $\vec r_2$ alone.
Thus, we can perform an average over the directions of $\vec r_2$. 
This translates into carrying out the sum over the $\mu$ component of the 
angular momentum $\lambda$.
We obtain the result
\begin{eqnarray}
   \rho^{\lambda}_{\alpha \beta \gamma \delta}(r_1,r_2,\theta_{12}) 
   & = & 
     \rho^{\lambda}_{\alpha \gamma}(r_1) \ 
     \rho^{\lambda}_{\delta \beta}(r_2) \ 
   P_{\lambda}(cos\theta_{12})
\end{eqnarray}

We now discuss the most dominant contributions. 
We again apply Eq.~(\ref{eq:Abar}) to evaluate the two-body density matrix, 
$ \langle \ {\mathbf a}^{\dag}_\alpha \ {\mathbf a}^{\dag}_\beta \ 
                {\mathbf a}_\delta \ {\mathbf a}_\gamma \ 
  \rangle $.
With this, we get the ground state two-body density as
\begin{eqnarray}
   &&
   \rho(r_1,r_2,\theta_{12})  
   \nonumber \\ &&
   = \ 
   \langle 0 \, | \, \rho_2^{op} \, | \, 0 \rangle
   \ + \ 
   \langle 0 
       \, | \, \rho_2^{op} \tilde {\mathbf S}_2^{\dag} 
       \, | \, 0 
   \rangle
   \ + \ 
   \langle 0 
       \, | \, {\mathbf S}_2 \rho_2^{op} 
       \, | \, 0 
   \rangle
   \nonumber \\ &&
   \ + \ 
   \langle 0 
       \, | \, \left [ {\mathbf S}_2, \ \rho_2^{op} \right ] \ 
               \tilde {\mathbf S}_2^{\dag} 
       \, | \, 0 
   \rangle
   \ + \ 
   \langle 0 
       \, | \, \rho_2^{op} \tilde {\mathbf S}_1^{\dag} 
       \, | \, 0 
   \rangle
   \nonumber \\ &&
   \ + \ 
   \frac{1}{2} \ 
   \langle 0 \, | \, 
       \left [ 
              {\mathbf S}_2, \ \left[ {\mathbf S}_2, \ \rho_2^{op} \right ] 
       \right ]
   \tilde {\mathbf S}_2^{\dag} 
   \, | \, 0 \rangle
   + \cdots
   \>.
\end{eqnarray}

(a)Contributions from bare ground state:
\begin{eqnarray}
   &&
   \rho(r_1,r_2,\theta_{12})
   \nonumber \\ &&
   = \ 
   \sum_{h_1h_2} \ (2j_{h_1}+1) \, R_{h_1}^2(r_1) \ 
                   (2j_{h_2}+1) \, R_{h_2}^2(r_2)
\end{eqnarray}
This term has the structure $\rho_{hf}(1) \rho_{hf}(2)$. Here
$\rho_{hf}$ is the bare one-body density. 
We can include all higher order one-body density terms by adding
\begin{equation}
   \Delta \rho(r_1,r_2,\theta_{12})=\Delta \rho(r_1)\rho_{hf}(r_2)
   +\rho_{hf}(r_1)\Delta \rho(r_2)
   \> ,
\end{equation}
where $\Delta \rho(r)$ is the difference between the full one-body density
minus the bare one-body density.
In turn, in higher order terms we have to exclude those terms where
in the operator $\rho_{\alpha \beta \gamma \delta}$ 
$\alpha$ connects with $\gamma$ or $\beta$ connects with $\delta$. 
However, we still have to consider the exchange terms
where $\alpha$ connects with $\delta$ and $\beta$ connects with $\gamma$.

The exchange term is due to the Pauli correlations and results in
\begin{eqnarray}
   &&
   \Delta\rho(r_1,r_2,\theta_{12}) 
   \nonumber \\ && 
   \ = \ 
   - \ 
   \sum_{h_1 h_2} \ 
   \sum_{l} \ 
   \rho^{l}_{h_1 h_2}(r_1) \ \rho^{l}_{h_1 h_2}(r_2) \ 
   P_{l}(cos\theta_{12})
   \> .
\end{eqnarray}

(b) Contributions linear in the $2p2h$-amplitudes.
For the following terms we assume a summation over all orbits appearing
twice in the expression. The leading term that arises from the
correlations is

\begin{eqnarray}
   \lefteqn{
   \Delta\rho(r_1,r_2,\theta_{12}) = 
   \sum_{\lambda} \ 
   \left ( S^{\lambda}_{p_1 h_1 p_2 h_2} \ + \ Z^{\lambda}_{p_1 h_1 p_2 h_2}
   \right )
   }
   \nonumber \\ && \hspace{1in} 
   \times
   \rho^{\lambda}_{h_1p_1}(r_1)
   \rho^{\lambda}_{p_2h_2}(r_2)P_{\lambda}(cos\theta_{12})
   \> .
\end{eqnarray}

In our application we have included all terms up to second order. Aside
from the terms listed here, the next most important correction arises from
the second order term depending on ${\mathbf S}_2$ and ${\mathbf \tilde
S}_2^{\dag}$.

\begin{figure}
   \epsfxsize = 2.0in
   \centerline{\epsfbox{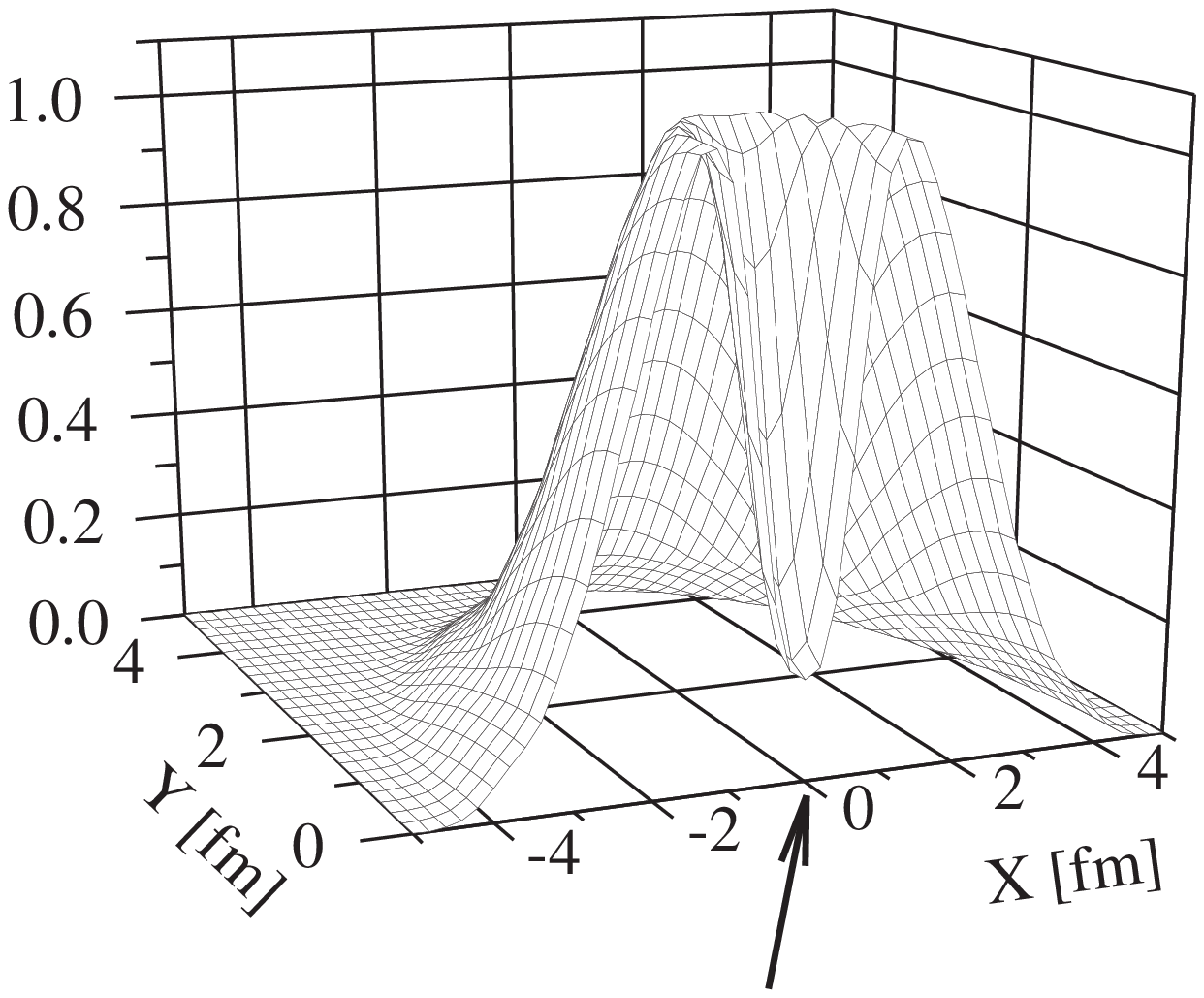}}
   \caption{The p-p two-body density
            for the Argonne $v_{18}$ potential: 
            the case when the first proton is located at $x_1 = 0.0 fm$}
   \label{fig:twoden-pp_0}
\end{figure}

\begin{figure}
   \epsfxsize = 2.0in
   \centerline{\epsfbox{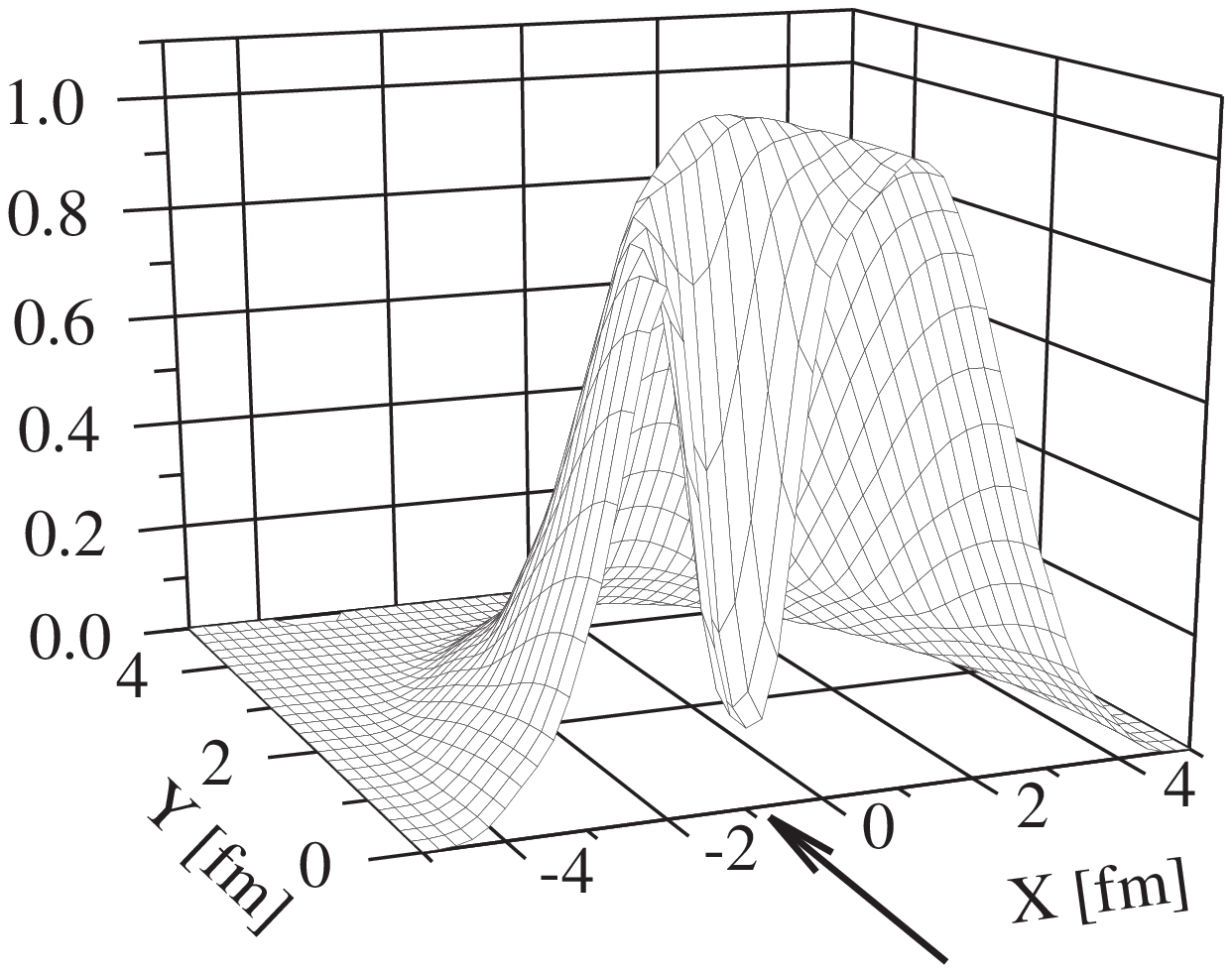}}
   \caption{Same as Fig.~\ref{fig:twoden-pp_0} 
            with the first proton located at $x_1 = 1.0 fm$}
   \label{fig:twoden-pp_1}
\end{figure}

\begin{figure}
   \epsfxsize = 2.0in
   \centerline{\epsfbox{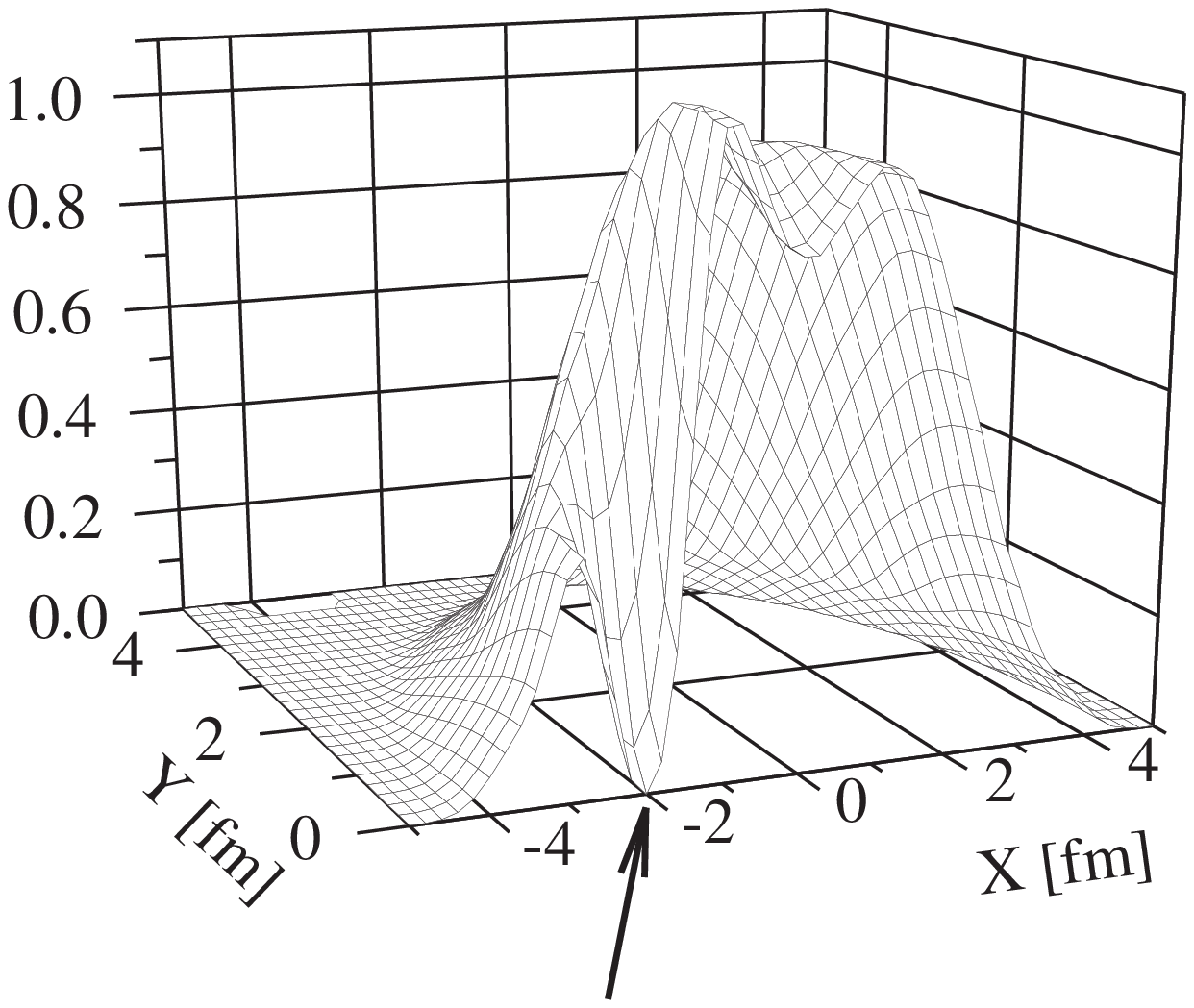}}
   \caption{Same as Fig.~\ref{fig:twoden-pp_0} 
            with the first proton located at $x_1 = 2.0 fm$}
   \label{fig:twoden-pp_2}
\end{figure}

\begin{figure}
   \epsfxsize = 2.0in
   \centerline{\epsfbox{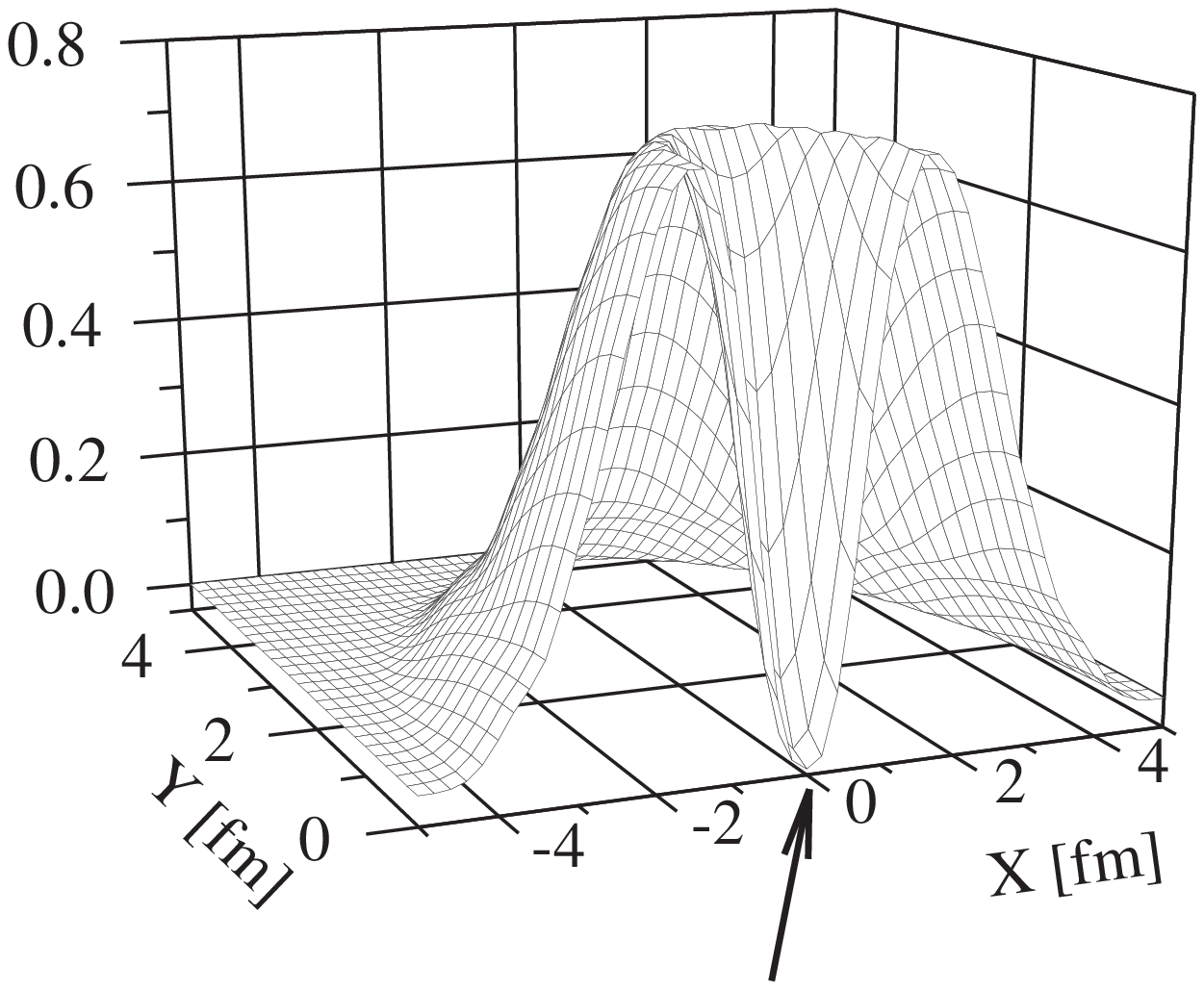}}
   \caption{The p-n two-body density
            for the Argonne $v_{18}$ potential:
            the case when the proton is located at $x_1 = 0.0 fm$}
   \label{fig:twoden-pn_0}
\end{figure}

\begin{figure}
   \epsfxsize = 2.0in
   \centerline{\epsfbox{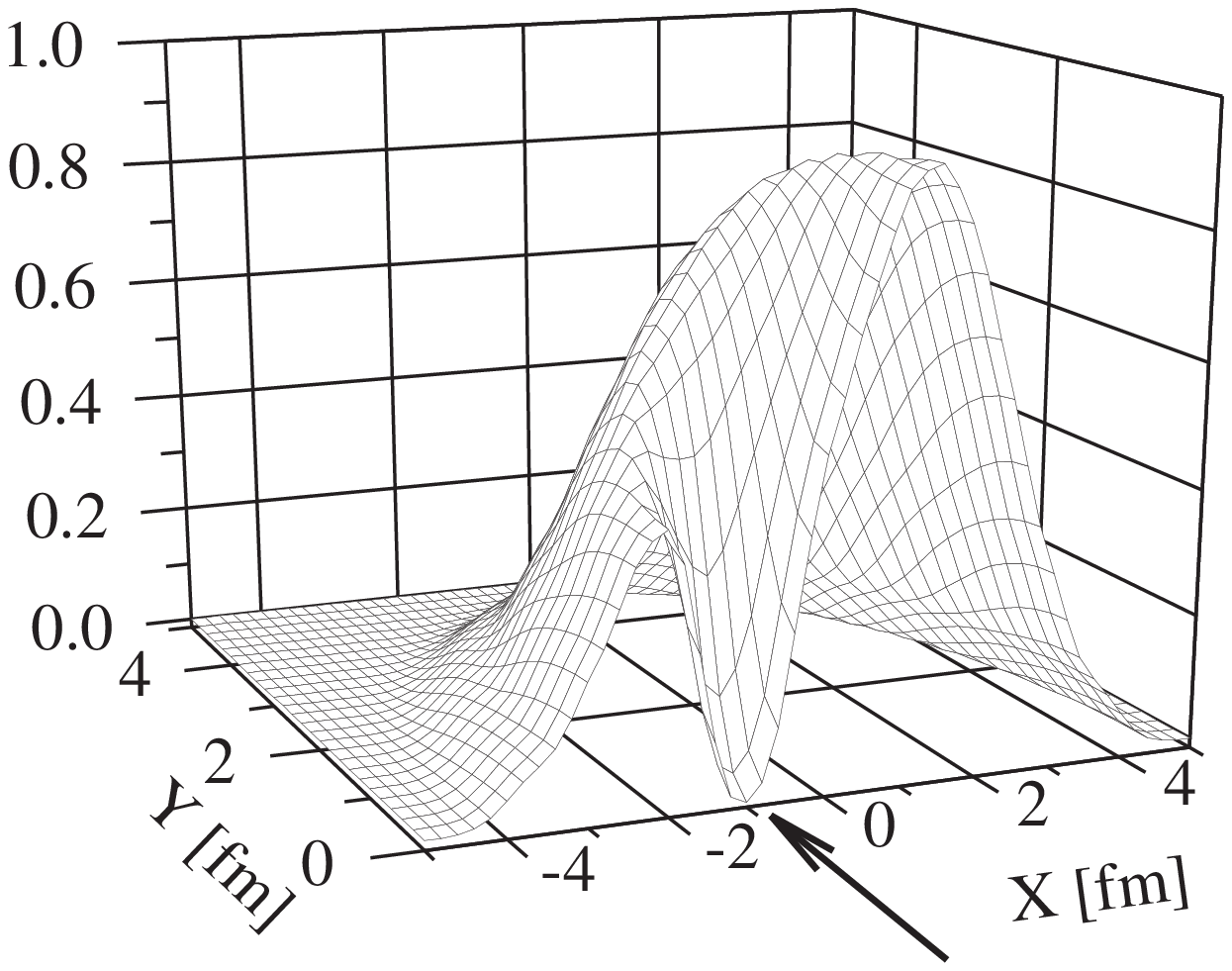}}
   \caption{Same as Fig.~\ref{fig:twoden-pn_0} 
            with the proton located at $x_1 = 1.0 fm$}
   \label{fig:twoden-pn_1}
\end{figure}

\begin{figure}
   \epsfxsize = 2.0in
   \centerline{\epsfbox{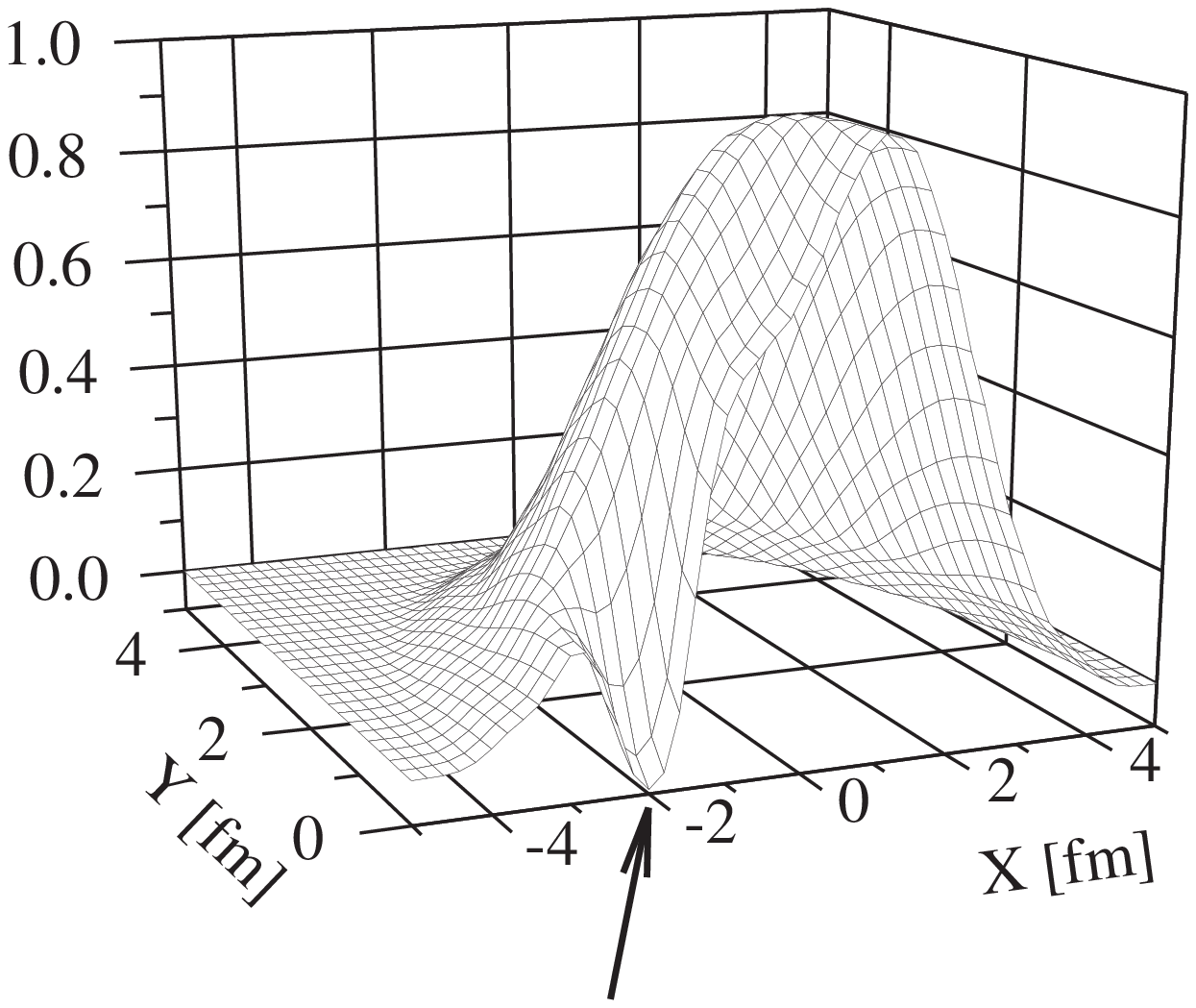}}
   \caption{Same as Fig.~\ref{fig:twoden-pn_0} 
            with the proton located at $x_1 = 2.0 fm$}
   \label{fig:twoden-pn_2}
\end{figure}

The two-body density represents the probability of finding one nucleon at 
$\vec r_1$ and one nucleon at $\vec r_2$. We can divide this by the
probability of finding the first nucleon at $\vec r_1$. The remaining
density represents the probability of finding a second nucleon at
$\vec r_2$ if the first nucleon is at $\vec r_1$. If both nucleons
are protons, this density is normalised to a total integral of (Z-1).
Figs.~\ref{fig:twoden-pp_0} to~\ref{fig:twoden-pn_2} 
show these densities as a function of $\vec r_2$ for 
various positions of $\vec r_1$. We have made no attempt to correct
these for the residual center-of-mass motion of the nucleus. 
The densities show the effects of the short range repulsion:
they exhibit a deep hole where the first nucleon is located.
The two-body densities also show that for large distances
the long-range aspect of the ground state nuclear correlations,
usually thought to be related with the surface deformation modes,
has a significant contribution:
when the first nucleon is located closer to the nuclear surface,
we observe an enhancement of the density at the symmetrically-opposite position.
The picture of a two-body density obtained as the revolution
of the spherically symmetric one-body density,
with a Gaussian-like distribution centered at the location of the
first nucleon scooped out of it, is definitely insufficient.

\section{Conclusions}

We use the coupled cluster expansion ($\exp(S)$ method) to solve the many-body
Schr\"odinger equation in configuration space.
While the coupled cluster expansion is exact if carried out to all orders,
the present results are obtained with truncations.
As we have retained only terms up to second order in $1/\epsilon$ the contributions
from ${\mathbf S}_5$ have no effect on ${\mathbf S}_2$.
In turn, the effects of any higher $n$ correlations are only of the order
as the higher order terms left out anyway.
Thus, in essence we have replaced the truncations in ${\mathbf S}_n$
by the more relevant truncation in $1/\epsilon$.

We have shown that it is possible to choose large enough configuration
spaces for the complete and self-consistent
calculation of the ground state correlations inside a finite nucleus. 
This calculation makes no artificial separation between
``short range'' and ``long range'' correlations. In fact, the two-body
density shows that the correlation function in the surface region of
the nucleus has strong contributions from the surface deformation modes.
It is largely these modes that cause the strong deoccupation
of orbits close to the Fermi surface.
The occupation of the 2$s_{1/2}$ and the 1$d_{5/2}$ proton orbits
calculated is consistent with the values suggested by the experiments.

Our efforts are currently directed in two directions. 
First, we intend to apply the procedure described in this paper 
to the particular case of a more realistic interaction, 
namely the Argonne $v_{18}$ potential~\cite{ref:WiringaETal} 
together with a phenomenological three-nucleon interaction~\cite{ref:tnipot}. 
This should not only result in a better description of the 
$^{16}$O observables, but a breakdown of the contributions 
to the binding from the two- and the three-body interactions 
could be inferred also.
Secondly, we shall use the equation of motion technique to calculate 
excited states of the $^{16}$O nucleus, 
and the ground state and excited states of the neighboring nuclei 
(e.g. $^{15}$N). 
We hope to be able to present our findings in the near future.

\section*{acknowledgements}

This work was supported by the U.S. Department of Energy 
(DE-FG02-87ER-40371).
Original calculations were carried out on a HP-9000/735 workstation 
at the Research Computing Center, 
and a dual-processor 200 MHz Pentium Pro PC at the 
Nuclear Physics Group of the University of New Hampshire.
The authors gratefully acknowledge the help of Mike Strayer and David Dean 
in letting us use their computing resources for 
the mapping of the binding energy with respect to the size 
of the configuration space. These calculations were carried out on a 
180 MHz R10000 Silicon Graphics Workstation at the 
Computational and Theoretical Physics Section ORNL.
The authors gratefully acknowledge John Dawson for his 
permanent support in every step of this project.
The authors also acknowledge helpful conversations with
Vijay~Pandharipande and Robert~Wiringa, 
who also supplied the Fortran subroutines to calculate the radial shape  
of the $NN$ interaction.


\newpage

%
%

\begin{table}
   \caption{Energy expectation values, charge radii, 
            and proton orbits occupation probabilities, 
            for the Argonne $v_8$, $v_{14}$ and $v_{18}$ potentials, respectively.}
   \begin{tabular}{lccccc}
      \\
      Potential & E       & rms   & 1$d_{5/2}$ & 2$s_{1/2}$ \\ 
                & [MeV/nucleon]        
                          & [fm]  & [\%]       & [\%]       \\
      \\
      \tableline \\
      $v$8      & - 7.0  & 2.81   & 3.68       & 4.09       \\ 
      $v$14     & - 6.1  & 2.86   & 3.33       & 3.99       \\
      $v$18     & - 5.9  & 2.81   & 2.58       & 2.75       \\
      \\
      expt.     & - 8.0   & 2.73  & 2.17       & 1.78       \\
                &         & $\pm$ 0.025 & $\pm$ 0.12 & $\pm$ 0.36 \\
      \\ 
   \end{tabular}
\label{tab:erms_occ_2b}
\end{table}

%
%



\end{document}